\RequirePackage[2020-02-02]{latexrelease}
\documentclass[12pt,showpacs,showkeys,amsmath,amssymb,nofootinbib]{revtex4}
\usepackage{amsmath,amsfonts,amsthm,amscd,amssymb,latexsym}
\usepackage{bm}
\usepackage{footnote}
\usepackage{dcolumn}
\usepackage{graphicx}
\usepackage{epstopdf}
\usepackage{color}
\usepackage{epsf}
\usepackage{epsfig}
\usepackage{graphicx, epic, eepic, color}
\usepackage[colorlinks,citecolor=blue,urlcolor=blue,linkcolor=blue]{hyperref}

\makeatletter
 \DeclareRobustCommand\ref{%
    \@ifstar\@refstar\T@ref
  }%
  \DeclareRobustCommand\pageref{%
    \@ifstar\@pagerefstar\T@pageref
  }%
 \makeatother

\begin{document}

\title{Review of the Gravitomagnetic Clock Effect}

\author{Lorenzo \surname{Iorio}$^{1}$}
\email{lorenzo.iorio@libero.it}
\author{Bahram \surname{Mashhoon}$^{2,3}$}
\email{mashhoonb@missouri.edu}

\affiliation{
$^1$Ministero dell'Istruzione e del Merito (M.I.U.R.) \\ Viale Unit\`a di Italia 68, I-70125, Bari (BA), Italy\\
$^2$Department of Physics and Astronomy, University of Missouri, Columbia, Missouri 65211, USA\\
$^3$School of Astronomy, Institute for Research in Fundamental
Sciences (IPM), Tehran 19395-5531, Iran\\
}

\date{\today}

\begin{abstract}
ABSTRACT:
The general relativistic gravitomagnetic clock effect, in its simplest form, consists of the non-vanishing difference in the orbital periods  of two counter-orbiting objects moving in opposite directions along circular orbits lying in the equatorial plane of a central rotating source.
We briefly review both the theoretical and observational aspects of such an intriguing consequence of Einstein's theory of gravitation.
\end{abstract}


\maketitle

\section{Introduction}

In the weak-field and slow-motion approximation, Einstein's \emph{general theory of relativity}  (GTR) implies that mass-energy currents generate gravitomagnetic fields, in close analogy with electrodynamics~\cite{Einstein, Thirring1, Thirring2,  L-Th, Mashhoon:1984fj}.
The  paradigm of gravitoelectromagnetism\footnotemark{} \footnotetext{For an historical overview, see Section IV of \cite{gem92}, and references therein. See also \textcolor{black}{the webpage} \url{http://www.phy.olemiss.edu/~luca/Topics/grav/gravitomagnetism.html} \textcolor{black}{maintained by Luca Bombelli} on the Internet. \textcolor{black}{Accessed 9th December 2023.}} (GEM)
\cite{1958NCim...10..318C, IRE58, IRE61, PTey1, PTey2, Thorne86, 1986hmac.book..103T, 1988nznf.conf..573T, 1991AmJPh..59..421H,
1992AnPhy.215....1J, gem92, gem98, 2001rfg..conf..121M, Clark:2000ff, 2001rsgc.book.....R, Mashhoon:2003ax, 2008PhRvD..78b4021C,
2014GReGr..46.1792C, Mashhoon:2019jkq, 2021Univ....7..388C, Costa:2019loe, 2021Univ....7..451R, Bini:2021gdb}
encompasses a series of phenomena affecting orbiting test particles, precessing gyroscopes, moving clocks and atoms, and propagating electromagnetic waves \cite{1977PhRvD..15.2047B, 1986SvPhU..29..215D, Tarta, Tartalight, Herrera:2001xk, 2002EL.....60..167T, 2002NCimB.117..743R, 2004GReGr..36.2223S, 2009SSRv..148...37S}.
In the case of an isolated rotating material body, the source of its stationary gravitomagnetic field turns out to be its proper angular momentum $\textbf{J}$.
The gravitomagnetic spin precessions \cite{Pugh59, Schiff60} of four spaceborne gyroscopes have recently been measured in the field of the Earth by means of the dedicated Gravity Probe B (GP-B) space experiment \cite{Francis1, Francis2}. For other ongoing or proposed tests of relativistic gravitomagnetism in our solar system, see, e.g., \cite{2011Ap&SS.331..351I, 2013CEJPh..11..531R}, and the references therein.\\

Among other things, the presence of the gravitomagnetic field alters the temporal structure of the aforementioned stationary source of mass $M$ and angular momentum $J$. A prominent illustration of this circumstance is provided by the gravitomagnetic clock effect (GCE) \cite{CoMa, Mashhoon:1997, Mashhoon:1998nn, Mashhoon:1999nr, Mashhoon:1998fj, You98, Tar2, Iorio:2001cb}.
It consists of the fact that, for a pair of free test particles moving in opposite senses about the central spinning body along identical equatorial and circular orbits, the gravitomagnetic field modifies their otherwise Keplerian orbital periods $T_0$ by small equal and opposite corrections $\delta T^{\pm}_{\mathrm{gvm}} = \pm \,2 \pi J/(Mc^2)$, where the plus (minus) sign refers to prograde (retrograde) motion. The fact that prograde motion takes longer than retrograde motion is in contradiction with the Machian notion of inertia.

The GCE is the orbital counterpart of the gravitational analogue of the Zeeman effect affecting electromagnetic waves propagating in the exterior gravitomagnetic field of a spinning body. In this case, the spectral line of an atom with frequency $\nu_0$, emitted in the stationary exterior field of a rotating body and received by a distant observer, splits into two components with opposite circular polarization and with frequencies $\nu_0\pm \delta\nu_\mathrm{gvm}$, where the plus (minus) sign refers to positive (negative) helicity radiation \cite{Zel65, book, Guts73}. For instance, in the case of electromagnetic waves propagating along the axis of rotation of the source, $\delta\nu_\mathrm{gvm} = G J/(\pi c^2 r^3)$, where $r$ is the radial distance of the observer from the source. This phenomenon is directly related to the gravitational Faraday rotation, also known as the Skrotskii effect, which involves the gravitomagnetic rotation of the plane of linear polarization of electromagnetic waves propagating in the exterior gravitational field of a rotating source~\cite{Skrotskii}.

In the exploration of GCE, one often considers counter-orbiting spaceborne clocks on satellites. The adoption of two counter-orbiting satellites in the field of the Earth was proposed for the first time in \cite{vanpa1, vanpa2} to measure the Lense--Thirring orbital precession \cite{L-Th}, which is another gravitomagnetic effect due to the angular momentum of a rotating body. A form of GCE for traveling electromagnetic waves was considered in \cite{Tarta,Tartalight}.

This review is organized as follows.
In Section\,\ref{rudi}, the general features of the GCE are presented. Section\,\ref{mach} elucidates the connection between the GCE and Mach's principle.
A derivation of GCE within the GEM framework is described in Section\,\ref{sec:nexus}.
Extensions and generalizations of the GCE are dealt with in Section\,\ref{exte}. Observational aspects of the GCE are the subject of Section\,\ref{obse}.
Section\,\ref{summ} offers a brief discussion of the presented results.


\section{Rudimentary GCE}\label{rudi}

In GTR, the gravitational field is described by the Riemannian curvature of the spacetime manifold with metric
\begin{equation}\label{I1}
\left(ds\right)^2 = g_{\mu \nu} \,dx^\mu \,dx^\nu\,.
\end{equation}
Here, the signature of the metric is +2, and latin and greek indices run from 1 to 3 and from 0 to 3, respectively. We employ units such that the speed of light in vacuum $c$ and Newton's gravitational constant $G$ are set equal to unity, i.e. $c = G = 1$, unless specified otherwise. Furthermore, we deal with spacetime coordinate systems $x^\mu$ that are admissible. We consider the class of preferred observers that are all spatially at rest in this gravitational field. According to the hypothesis of locality, a preferred observer measures proper time $\tau_{\rm S}$ given by
\begin{equation}\label{I2}
\tau_{\rm S}  = \int_0^t  (-g_{00})^{1/2} dx^0\,,
\end{equation}
where  $x^0 = ct$ and we assume $\tau_{\rm S} = 0$ at coordinate time $t = 0$.

Let us now imagine preferred observers that are spatially at rest on the equatorial plane of the exterior  Kerr spacetime \cite{Kerr63, TK15} whose gravitational source is endowed with mass $M$ and angular momentum $J$. The Kerr source is rotating in the positive sense about the $z$ direction.  Furthermore, the spacetime is stationary and axially symmetric, with timelike and azimuthal Killing vector fields $\partial_t$ and $\partial_\varphi$, respectively. The Kerr metric is given in the standard Boyer--Lindquist coordinates $(t,r,\theta,\varphi)$ by~\cite{BL67, Chandra}
\begin{equation}\label{I3}
\left(ds\right)_{\rm K}^2=-\left(dt\right)^2+\frac{\Sigma}{\Delta}\left(dr\right)^2+\Sigma\, \left(d\theta\right)^2 +(r^2+a^2)\sin^2\theta\, \left(d\varphi\right)^2+\frac{2Mr}{\Sigma}(dt-a\sin^2\theta\, d\varphi)^2\,,
\end{equation}
where $a := J/M$ is the specific angular momentum of the source, also known as the Kerr parameter. Here,
\begin{equation}\label{I4}
\Sigma := r^2 + a^2\cos^2\theta\,,\qquad \Delta := r^2 - 2Mr + a^2\,.
\end{equation}
The gravitational potentials for a rotating source in GTR can be naturally interpreted in terms of a dimensionless gravitoelectric potential $GM/(c^2 r)$ and a gravitomagnetic potential $GJ/(c^3 r^2)$. We are interested in the exterior Kerr spacetime with $r \gg M$. The simple analogy with electromagnetism leads to GEM, which we will use throughout this review.

We are interested in future-directed timelike and stable circular geodesic orbits in the equatorial plane of the Kerr source. The geodesic equation is given by
\begin{equation}\label{I5}
\frac{d^2x^\alpha}{d\tau^2} + \Gamma^{\alpha}_{\mu \nu} \,\frac{dx^\mu}{d\tau}\,\frac{dx^\nu}{d\tau} = 0\,,
\end{equation}
where $\tau$ is the proper time and
\begin{equation}\label{I6}
\Gamma^\mu_{\alpha \beta}= \frac{1}{2} g^{\mu \nu} (g_{\nu \alpha,\beta}+g_{\nu \beta,\alpha}-g_{\alpha \beta,\nu})\,.
\end{equation}
The radial component of the geodesic equation for constant radial coordinate $x^1 = r =r_0$ and $\theta = \pi/2$ can be written as
\begin{equation}\label{I6a}
\Gamma^{1}_{0 0} \,\left(\frac{dt}{d\varphi}\right)^2 + 2\,\Gamma^{1}_{0 3} \,\frac{dt}{d\varphi} + \Gamma^{1}_{3 3} = 0\,,
\end{equation}
which simplifies considerably due to the manageable form of the relevant Christoffel symbols in this case, namely,
\begin{equation}\label{I6b}
g_{00, 1} \,\left(\frac{dt}{d\varphi}\right)^2 + 2\,g_{03, 1} \,\frac{dt}{d\varphi} + g_{3 3, 1} = 0\,.
\end{equation}

 The result, for Kerr circular equatorial orbits, is
\begin{equation}\label{I7}
\left(\frac{dt}{d\varphi} - a \right)^2 = \frac{r_0^3}{M}\,.
\end{equation}
Let $\omega_0$ be the Keplerian frequency of the orbit
\begin{equation}\label{I8}
\omega_0 = \left(\frac{M}{r_0^3}\right)^{1/2},
\end{equation}
so that the Keplerian orbital period is
\begin{equation}
T_0 = \frac{2\pi}{\omega_0};
\end{equation}
then, Equation~\eqref{I7} implies $dt/d\varphi = \pm \,\omega_0^{-1} + a$; that is,
\begin{equation}\label{I9}
\frac{dt}{d\varphi} = \pm\, \frac{1}{\omega_0} (1 \pm a \omega_0)\,,
\end{equation}
where the upper (lower) sign indicates a co-rotating (counter-rotating) orbit with respect to the rotation of the central body. Integrating this relation with respect to the azimuthal coordinate $\varphi$ over $(0, 2\pi)$ for the co-rotating orbit and over  $(0, -2\pi)$ for the counter-rotating orbit,
we find\footnotemark{} \footnotetext{To the authors' knowledge, the gravitomagnetic correction $2\pi a$ of the Keplerian orbital period for a prograde orbit was calculated, for the first time, on p. 91 of \cite{Vladi87}.}
\begin{equation}\label{I10}
t_{\pm} = T_0 + \delta T^{\pm}_\mathrm{gvm} = \frac{2\pi}{\omega_0} \pm 2\pi a\,
\end{equation}
and
\begin{equation}\label{I11}
 t_{+} - t_{-} = 4 \pi a\,.
\end{equation}
See \cite{CoMa, Mashhoon:1998nn, You98, Tar2, Iorio:2001cb} for various approaches in deriving Equation~\eqref{I11}.
According to the preferred observer located at $r_0$,  $t_{+}$ ($t_{-}$) is the orbital period for the co-rotating (counter-rotating) circular geodesic path in terms of the temporal coordinate.  For $a = 0$, the coordinate periods of the two orbits are equal and Keplerian.  Using Equation~\eqref{I2}, the corresponding  proper orbital periods according to the preferred observer located at $r_0$ are $\tau_{{\rm S}\pm} = (1-2M/r_0)^{1/2} \,t_{\pm}$, since observables in GTR are scalar invariants.  For the Earth\footnotemark{}\footnotetext{According to \cite{iers10}, Earth's angular momentum per unit mass amounts to $J_\oplus/M_\oplus\simeq 9\times 10^8\,\mathrm{m^2\,s^{-1}}$; thus, the terrestrial angular momentum is $J_\oplus\simeq 5.85\times 10^{33}\,\mathrm{kg\,m^2\,s^{-1}}$.}, $4\pi a_{\oplus} \approx 1.37 \times 10^{-7}$ s, while $GM_{\oplus}/(c^2 r_{\oplus}) \approx 7 \times 10^{-10}$; therefore, $\tau_{{\rm S} +} - \tau_{{\rm S} -} \approx t_{+} - t_{-} = 4 \pi a$.

The topological nature of Equation~\eqref{I11} should be noted\footnotemark{} \footnotetext{See also the discussion on p. 92 of \cite{Vladi87}.}: the difference in the orbital periods is independent of the coordinate radius of the orbit ($r$) and depends only on the angular momentum per unit mass of the central Kerr source. We have here a gravitomagnetic analogue of the Aharanov--Bohm effect~\cite{AB}.  Another significant feature of this GCE is its independence from the gravitational coupling constant $G$.

Let us briefly digress here and mention that stable circular equatorial geodesic orbits exist in Kerr spacetime for a test particle from $r_0$ near  $\infty$ down to the solutions of~\cite{Chandra}
\begin{equation}\label{I12}
1-\frac{6M}{r_0}\pm 8 a \omega_0-3\frac{a^2}{r_0^2}=0\,.
\end{equation}
As before,  the upper (lower) sign refers to orbits where the test particle rotates in the same (opposite) sense as the source. For $a = 0$, $r_0 = 6 M$. When the orbital radius decreases below the values given by Equation~\eqref{I12},
there are then unstable circular orbits all the way down to the null circular geodesic orbits given by
\begin{equation}\label{I13}
1-\frac{3M}{r_0}\pm 2 a \omega_0 = 0\,.
\end{equation}

It is interesting to imagine standard clocks moving on the circular geodesic orbits under consideration here. What is the total amount of proper time that a standard clock registers after it completes the orbit according to the preferred observer? The metric of Kerr spacetime~\eqref{I3} implies
\begin{equation}\label{I14}
\left(d\tau\right)^2 = \left(1-\frac{2M}{r}\right) \left(dt\right)^2 + 4\, \frac{Ma}{r} dt\, d\varphi - (r^2 + a^2 + 2Ma^2/r) \left(d\varphi\right)^2\,.
\end{equation}
Using Equation~\eqref{I9}, it is straightforward to show, after some algebra, that
\begin{equation}\label{I15}
\frac{d\tau}{d\varphi} = \pm\,\frac{1}{\omega_0} \left(1 -3\frac{M}{r_0} \pm\, 2 a \omega_0\right)^{1/2}\,.
\end{equation}
As in Equation~\eqref{I10}, we find upon integration over the azimuthal coordinate
\begin{equation}\label{I16}
\tau_{\pm} = \frac{2\pi}{\omega_0} \left(1 -3\frac{M}{r_0} \pm\, 2 a \omega_0\right)^{1/2}\,.
\end{equation}
Assuming that $a \omega_0 \ll 1$ and working to first order in this quantity, we obtain
\begin{equation}\label{I17}
\tau_{\pm} = \frac{2\pi}{\omega_0} \left(1 -3\frac{M}{r_0}\right)^{1/2}  \pm\, \frac{2\pi a}{\left(1 -3\frac{M}{r_0}\right)^{1/2}}\,.
\end{equation}
Therefore, for $r_0 \gg M$, we have
\begin{equation}\label{I18}
\tau_{+}  - \tau_{-} \approx 4\pi a\,.
\end{equation}
As determined by the preferred observer,  standard spaceborne clocks on  circular equatorial geodesic orbits about the Kerr source with the same coordinate radius $r$ register  longer proper times to complete  co-rotating orbits than  counter-rotating orbits and the difference in the corresponding proper times is approximately $4 \pi J/(Mc^2)$, which  is independent of $r_0$ and $G$.

For the Earth, $4\pi a_{\oplus} \approx 1.37 \times 10^{-7}$ s, which is per se easily measurable. Though Eqs.~\eqref{I11} and~\eqref{I18} are, in principle, rather promising, their experimental verification appears intractable; in this connection, see also Section\,\ref{obse}. To see this, imagine two neighboring circular equatorial orbits around the Earth. Then,
\begin{equation}\label{I19}
\frac{\delta T_0}{T_0} = \frac{3}{2}\, \frac{\delta r}{r_0}\,.
\end{equation}
For near-Earth orbits with $\delta T_0 \approx 1.37 \times 10^{-7}$ s, we find $\delta r \approx 0.02$ cm. Submillimeter orbital control that would be needed for experimental measurement of the effects illustrated in Eqs.~\eqref{I11} and~\eqref{I18} appears to be beyond the present capabilities. Therefore, one should concentrate instead on the orbital periods given in Eqs.~\eqref{I10} and~\eqref{I17}; for orbits of different orbital radii $r$,  the contribution of the rotation of the source to the orbital period is $+2 \pi a$ for prograde orbits and $-2\pi a$ for retrograde orbits. For proper periods, the dependence of this effect on the orbital radius can be practically ignored; for instance, in the case of the Earth $GM_{\oplus}/(c^2 r_{\oplus}) \approx 7\times 10^{-10}$. Nevertheless, it appears  that the observation of the contribution of the angular momentum of the source to the orbital period is a daunting task on a par with the GP-B experiment, since the relative size of the effect, i.e. $2 \pi a/ T_0$, can be expressed as
\begin{equation}\label{I20}
\frac{2 \pi a}{T_0} = \frac{GJ/(c^2 r_0^3)}{\omega_0}\,,
\end{equation}
where $GJ/(c^2 r_0^3)$ is proportional to the magnitude of the gravitomagnetic precession frequency of the gyroscopes in the GP-B experiment~\cite{Francis1, Francis2} and is related to the gravitomagnetic field of the source~\cite{Bahram, You98, CC23}.

Imagine now that the observer is not spatially at rest and moves on a circle in the equatorial plane around the source. The orbital periods are affected by the motion of the observer and the resulting observer-dependent gravitomagnetic clock effects have been discussed in detail~\cite{Mashhoon:1998nn, BoSt, OSem, Bini:2000ge, Bini:2001kx, Maartens:2001nj}.

Finally, consider a thought experiment\footnotemark{} \footnotetext{In this connection, see the ``planetary gravitational Zeeman effect" of Mitskevich and Pulido Garcia~\cite{MitPG}.} involving two free standard clocks counter-revolving on the equatorial circular orbit of radius $r_0$. Their starting event is assumed to be at $\varphi=0$ and $t = 0$, i.e. at event $x^\mu = (t, r_0, \theta, \varphi)$ given by $(0, r_0, \pi/2, 0)$. Their first meeting is then at event $(t_1, r_0, \pi/2, \varphi_1)$, where
\begin{equation}\label{I21}
\varphi_1 = \pi(1-a\omega_0)\,, \qquad t_1 =\frac{1}{2} T_0 (1-a^2\omega_0^2)\,.
\end{equation}
Proceeding in this way, we find that their $n$th meeting point is at event $(t_n, r_0, \pi/2, \varphi_n)$, where
\begin{equation}\label{I22}
\varphi_n = n \pi(1-a\omega_0) ~~{\rm modulo}~~ 2 \pi\,, \qquad t_n =  \frac{1}{2} nT_0 (1-a^2\omega_0^2)\,.
\end{equation}
The proper times of the clocks at their $n$th meeting point is given by
\begin{equation}\label{I23}
\tau_n^{\pm} = \frac{1}{2} nT_0 (1\mp a\omega_0) \left(1 -3\frac{M}{r_0} \pm\, 2 a \omega_0\right)^{1/2}\,.
\end{equation}
From this result we can derive the observer-independent GCE  given to first order in $a/M$ by
\begin{equation}\label{I24}
\tau_n^{+} - \tau_n^{-} \approx  \frac{6 \pi n J}{[r_0(r_0-3M)]^{1/2}}\,.
\end{equation}
As the free clocks continue to meet, the diametrical line that joins the center of the circle of radius $r_0$ to a meeting point precesses in the opposite sense as the rotation of the source with a frequency of $\approx n\pi a \omega_0/\tau_n^{+}$. This quantity is proportional to $GJ/(c^2r_0^3)$ and corresponds to the precession frequency of an ideal free test gyroscope held at rest at radius $r_0$ in the equatorial plane of the Kerr source.


\section{GCE and Mach's Principle}\label{mach}

We have shown that free test particles on circular equatorial orbits about a Kerr source move slower (faster) on co-rotating (counter-rotating) orbits and take more (less) time to go around the source as a direct consequence of its rotation. Indeed, it is straightforward to demonstrate in the case of the Kerr source that the faster it rotates, the slower the prograde motion and the faster the retrograde motion. In the absence of the rotation of the source, the gravitomagnetic clock effect disappears for the static Schwarzschild source.  It appears that these results do not support Mach's principle \textcolor{black}{as postulated by Einstein. In fact},  we find in Einstein's book~\cite{Einstein} a discussion of Mach's ideas regarding inertia:

\emph{``But in the second place, the theory of relativity makes it appear probable that Mach was on the right road in his thought that inertia depends upon a mutual action of matter. For we shall show in the following that, according to our equations, inert masses do act upon each other in the sense of the relativity of inertia, even if only very feebly. What is to be expected along the line of Mach's thought?
1. The inertia of a body must increase when ponderable masses are piled up in its neighbourhood.
2. A body must experience an accelerating force when neighbouring masses are accelerated, and, in fact, the force must be in the same direction as that acceleration.
3. A rotating hollow body must generate inside of itself a `Coriolis field', which deflects moving bodies in the sense of the rotation, and a radial centrifugal field as well."}

Einstein's first point has been discussed in detail by Brans~\cite{Brans1, Brans2}. Adopting the modern geometric interpretation of GTR, Brans showed that the inertial mass of a free test particle in a gravitational field in an invariant. This  is now a completely resolved issue in GTR. The second point  is directly relevant to the gravitomagnetic clock effect and is in contradiction with the GCE~\cite{Mashhoon:1999nr}.  The third point has been extensively treated in the commentary on the Thirring--Lense work~\cite{Mashhoon:1984fj}.

In his influential book~\cite{Mach}, Mach rejected the Newtonian absolute acceleration and replaced it with acceleration relative to the masses in the universe. Einstein called this ``Mach's Principle"; moreover,  Einstein, following Mach,  assumed that inertia of matter is solely due to the presence of ambient masses (relativity of inertia). These ideas have contributed significantly to the development of general relativity; however, they are at present only of historical interest.  The current situation in GTR regarding acceleration is essentially similar to Newtonian physics, except that global inertial frames are now replaced by  local inertial frames. That is, inertial (d'Alembert) forces which appear in a laboratory that accelerates with respect to the local inertial frames, are not due to the gravitational influence of distant masses. Indeed,  inertial effects are not of gravitational origin; instead, the gravitational field of external masses would generate gravitational tidal effects in the laboratory~\cite{Mashhoon:1984fj, Lichtenegger:2004re, Mashhoon:2007qm, Mashhoon:2015nea}. \textcolor{black}{We have adopted Einstein's interpretation of Mach's ideas regarding inertia; however, there has been a diversity of opinion in connection with Mach's views. As a matter of fact}, Mach's profound and intriguing ideas have been explored by a number of authors; see, for instance,~\cite{Barbour:1995iu, Lichtenegger:2009gk, Mashhoon:2011qt, Barbour:2011ku, Mashhoon:2011kw, Glampedakis:2022fqu}.

Imagine an observer following a timelike world line $x^\mu (\eta)$, where $\eta$ is the observer's proper time. Let $u^\mu = dx^\mu/d\eta$ be the 4-velocity of the observer and $\mathcal{A}^\mu = Du^\mu/d\eta$ be its 4-acceleration. The observer's 4-velocity is a unit timelike vector, $u^\mu u_\mu = -1$, which, upon covariant differentiation, implies $u_\mu \mathcal{A}^\mu = 0$; hence, $\mathcal{A}^\mu$ is a spacelike 4-vector with magnitude $\mathbb{A}$, namely,
\begin{equation}\label{M1}
 \mathcal{A}_\mu \mathcal{A}^\mu := \mathbb{A}^2\,, \qquad \mathbb{A} \ge 0\,.
\end{equation}
That is, the observer's 4-acceleration is orthogonal to its timelike 4-velocity; therefore, the observer's 4-vector of translational acceleration $\mathcal{A}^\mu$ is spacelike. If the magnitude of translational acceleration vanishes, $\mathbb{A} = 0$, then the observer is free and follows a geodesic; otherwise, $\mathbb{A} > 0$ and the observer is accelerated as a consequence of being subjected to a non-gravitational force. This acceleration is independent of any system of coordinates and is in this sense absolute. In any local inertial frame established along the observer's world line, the non-gravitational force can be identified with the basic forces of nature that have been discovered in Minkowski spacetime.

The observer in general carries an orthonormal tetrad frame $e^\mu{}_{\hat {\alpha}}(\eta)$ for measurement purposes, where $e^\mu{}_{\hat 0}(\eta) = u^\mu$ and
\begin{equation}\label{M2}
 g_{\mu \nu} \, e^\mu{}_{\hat {\alpha}}\, e^\nu{}_{\hat {\beta}} = \eta_{\hat {\alpha} \hat  {\beta}}\,.
\end{equation}
Here, $\eta_{\alpha \beta}$ is the Minkowski metric tensor given by diag$(-1,1,1,1)$; moreover, we employ hatted indices to enumerate the tetrad axes in the local tangent space. The local manner in which the observer carries its tetrad frame along its world line defines the acceleration tensor $\Phi_{\hat {\alpha} \hat {\beta}}$, namely,
\begin{equation}\label{M3}
\frac{D e^\mu{}_{\hat {\alpha}}}{d\eta} =  \Phi_{\hat {\alpha}}{}^{\hat {\beta}} \, e^\mu{}_{\hat {\beta}}\,.
\end{equation}
The orthonormality condition~\eqref{M2} implies that the acceleration tensor is antisymmetric, $\Phi_{\hat {\alpha} \hat {\beta}}(\eta) = - \Phi_{\hat {\beta} \hat {\alpha}}(\eta)$. In simple analogy with electrodynamics, the acceleration tensor can be decomposed into its ``electric" and ``magnetic" components given by the locally measured translational acceleration and the locally measured angular velocity of the observer's local spatial frame relative to a nonrotating (i.e. Fermi--Walker transported) frame, respectively~\cite{Mashhoon:2003ax}. That is,
\begin{equation}\label{M4a}
 \Phi_{\hat 0 \hat i} = \mathcal{A}_{\hat i} = \mathcal{A}_\mu \, e^\mu{}_{\hat i}\,, \qquad  \Phi_{\hat i \hat j} = \epsilon_{\hat i \hat j \hat k}\, \Omega^{\hat k}\,,
\end{equation}
where $\Omega^\mu = \Omega^{\hat k}\,e^\mu{}_{\hat k}$. From $\Phi^{\mu \nu} =  \Phi^{\hat {\alpha} \hat {\beta}}\, e^\mu{}_{\hat {\alpha}}\, e^\mu{}_{\hat {\beta}}$, we find
\begin{equation}\label{M4b}
 \Phi^{\mu \nu} = \mathcal{A}^\mu\, u^\nu - \mathcal{A}^\nu \,u^\mu + \epsilon^{\mu \nu \rho \sigma}\, u_\rho \Omega_\sigma\,,
\end{equation}
where the alternating tensor is such that $\epsilon_{\hat 0 \hat 1 \hat 2 \hat 3} = 1$.

In classical physics,  inertial forces act on masses whose motions are described relative to noninertial frames of reference. They consist of d'Alembert's force in the case of the rectilinear acceleration of the frame,  as well as the Coriolis force, the centrifugal force and Euler's force in the case of the rotation of the frame. An inertial force acting on a body is naturally  proportional to its inertial mass as a consequence of Newton's laws of motion.  The inertial mass of a particle is equal to its gravitational mass in accordance with the principle of equivalence of inertial and gravitational masses. This circumstance appeared to suggest the notion that inertial forces could be of gravitational origin, a concept that must be rejected according to the modern geometric interpretation of GTR.

The standard results of GTR presented here have the physical implication that an observer confined to a small window-less laboratory that moves in a gravitational field can determine by means of purely local experiments within the laboratory whether the laboratory is being subjected to translational acceleration and rotation. By monitoring the motion of free test particles relative to the walls of the laboratory, the translational acceleration of the laboratory can be ascertained. Furthermore, the proper rotation of the laboratory can be determined from the precession of  Foucault's pendulum within the laboratory.

In quantum theory, the inertial properties of a particle are governed by its mass and spin, which characterize the irreducible unitary representations of the inhomogeneous Lorentz group~\cite{Wigner}. The inertial properties of  mass are well known from classical physics; moreover, the moment of inertia for rotational motion is the analogue of inertial mass for linear motion. On the other hand, the inertial properties of intrinsic spin have been the subject of recent investigations~\cite{Mashhoon:2015nea, DSH, DDSH, DDKWLSH, Yu:2022vjn}.  The inertia of intrinsic spin, which is completely independent of the inertial mass, elevates the issue of the origin of inertia to  the quantum domain. Regarding GTR,  mass-energy is essential in generating classical gravitational fields, while intrinsic spin couples to the gravitomagnetic field of a rotating source. A consequence of this spin-gravitomagnetic field coupling is the violation of the universality of free fall; for instance, a free neutron with spin up will in general fall differently in the gravitational field of the Earth than a free neutron with spin down~\cite{BaMa, Emelyanov:2022weg}. For a recent treatment of spin-gravity coupling, see~\cite{Mashhoon:2023idh}.


\section{GEM-GCE Nexus}\label{sec:nexus}

It proves useful to consider the derivation of GCE within the GEM framework~\cite{Iorio:2001cb}.  We start with the general linear approximation in GTR appropriate for the weak exterior gravitational field of a slowly rotating astronomical source  of mass $M$ and angular momentum $J$. For the sake of definiteness, we use mainly the notation and conventions of~\cite{2001rfg..conf..121M}. The spacetime metric is given by
\begin{equation}\label{N1}
(ds)_\mathrm{GEM}^2=-c^2\left(1-2\frac{\Phi_g}{c^2}\right)(dt)^2-\frac{4}{c}(\mathbf{A}_g \cdot d\mathbf{x}) dt+\left(1+2\frac{\Phi_g}{c^2}\right) \delta_{ij}dx^idx^j\,,
\end{equation}
where $\Phi_g$ is the gravitoelectric potential and $\mathbf{A}_g$ is the gravitomagnetic vector potential. All post-Newtonian terms of order $c^{-4}$ and smaller have been neglected in our GEM metric. Moreover, we assume that the source is stationary and $\nabla \cdot \mathbf{A}_g = 0$.  The GEM fields, defined by
\begin{equation}\label{N2}
\mathbf{E}_g := -\nabla \Phi_g\,, \qquad \mathbf{B}_g := \nabla \times \mathbf{A}_g\,,
\end{equation}
are, as expected, gauge invariant~\cite{2001rfg..conf..121M, Clark:2000ff}. Far from the source, we have the asymptotic relations
\begin{equation}\label{N3}
\Phi_g \sim \frac{GM}{r}\,,\qquad  \mathbf{A}_g \sim \frac{G\mathbf{J} \times \mathbf{x}}{c\, r^3}\,,
\end{equation}
where $r := |\mathbf{x}|$. Therefore,
\begin{equation}\label{N4}
\mathbf{E}_g \sim \frac{GM \mathbf{x}}{r^3}\,,\qquad \mathbf{B}_g \sim \frac{G}{c}\,\frac{3\,(\mathbf{J}\cdot \mathbf{x})\,\mathbf{x} - \mathbf{J}\,r^2}{r^5}\,.
\end{equation}

For the motion of a free test particle of mass $m \ll M$ in this gravitational field, we find to lowest order in $|\mathbf{v}|/ c \ll 1$, the force law
\begin{equation}\label{N5}
\mathbf{F} = -m \mathbf{E}_g  - 2 m \frac{\mathbf{v}}{c} \times \mathbf{B}_g\,,
\end{equation}
in close analogy with the Lorentz force of electrodynamics. Restricting this equation to a circular equatorial orbit of radius $r_0$, we find
\begin{equation}\label{N6}
\frac{v_{\pm}^2}{r_0} = \frac{GM}{r_0^2}\mp \frac{2 G J v_{\pm}}{c^2 r_0^3}\,,
\end{equation}
where the circular velocity has magnitude $v_{\pm} = r_0\, \omega_{\pm} = r_0 \,d\varphi_{\pm}/dt >0$ such that the plus (minus) sign indicates prograde (retrograde) motion. Equation~\eqref{N6} can be reduced to Equation~\eqref{I9} if  $J^2\omega_0^2/(M^2c^4)$ can be neglected in comparison with unity; in this way, we find  that $t_{\pm} \approx T_0 \pm 2\pi J/(Mc^2)$. As in Section\,\ref{rudi}, we can express this result in terms of the proper time of an observer that is spatially at rest in spacetime or with reference to the proper times of spaceborne clocks. In any case, GEM provides a more intuitive approach to GCE.

Are there other phenomena of a similar nature as GCE that can be elucidated within the GEM framework? Consider the Landau--Lifshitz pseudotensor $t_{\mu \nu}$ for the stationary slowly rotating source under consideration here. We find that the dominant contributions to $t^{\mu \nu}$ are given by~\cite{2001rfg..conf..121M}
\begin{equation}\label{N7}
t^{00} = -\frac{7}{8\pi} \frac{G M^2}{r^4}\,, \qquad (t^{0i}) = -\frac{1}{2\pi} \frac{G M J}{r^5}\sin \theta\, \boldsymbol{\hat{\varphi}}\,, \qquad t^{ij} =  \frac{1}{4\pi} \frac{G M^2}{r^6} x^i x^j\,.
\end{equation}
Here, $\boldsymbol{\hat{\varphi}}$ is a unit vector in the azimuthal direction. There is therefore an energy flow around the rotating mass with flow velocity $v_g^i = t^{0i}/t^{00}$; that is,
\begin{equation}\label{N8}
\mathbf{v}_g = \alpha\, \frac{J}{M\,r}\sin \theta\, \boldsymbol{\hat{\varphi}}\,,
\end{equation}
where $\alpha = 4/7$ for the Landau--Lifshitz pseudotensor. This form of flow velocity can be obtained using other definitions for the energy-momentum of the gravitational field in the case of the stationary exterior of a rotating mass, except that the coefficient $\alpha$ assumes other positive numerical values~\cite{2001rfg..conf..121M, Mashhoon:2019jkq}; that is, the sense of steady circular flow is always the same as the rotation of the source.

We note that the flow is divergence free but has vorticity, namely,
\begin{equation}\label{N9}
\nabla \cdot \mathbf{v}_g = 0\,, \qquad    \nabla \times  \mathbf{v}_g = \boldsymbol{\varpi}_g = 2 \alpha\, \frac{J}{M\,r^2}\cos \theta\, \mathbf{\hat{r}}\,,
\end{equation}
where $\mathbf{\hat{r}}$ is a unit vector in the radial direction. The vorticity $\boldsymbol{\varpi}_g$ is maximum along the rotation axis and vanishes in the equatorial plane. The flow streamlines are circles in planes parallel to the equatorial plane. The gravitational stream function is given by $\boldsymbol{\Psi}_g$, where $\mathbf{v}_g = \nabla \times \boldsymbol{\Psi}_g$. The gravitational stream function in our case takes the form
\begin{equation}\label{N10}
\boldsymbol{\Psi}_g = \alpha\, \frac{J}{M}\cos \theta\, \mathbf{\hat{r}}\,, \qquad   \nabla \cdot \boldsymbol{\Psi}_g = 2\alpha\, \frac{J}{M\,r}\cos \theta\,.
\end{equation}
The stream function depends only upon $J/M$ and its magnitude is independent of the radial coordinate, which is reminiscent of the GCE. Similarly, the circulation of the flow velocity is the line integral of  $\mathbf{v}_g$ along a streamline, namely,
\begin{equation}\label{N11}
\mathcal{C}_g = \oint \mathbf{v}_g \cdot d\boldsymbol{\ell}\,
\end{equation}
and is given in our case by
\begin{equation}\label{N12}
\mathcal{C}_g =  2 \pi \alpha\, \frac{J}{M}\sin^2 \theta\,,
\end{equation}
which is again independent of the radial distance $r$ to the source. The circulation vanishes along the axis of rotation, is maximum on the equatorial plane and depends only on the specific angular momentum of the source just like the GCE. However, it has not been possible to establish a direct physical connection between the flow of gravitational energy around a rotating mass and the gravitomagnetic clock effect.


\section{Extensions and Generalizations of GCE}\label{exte}

The treatment of GCE in connection with circular equatorial orbits about Kerr spacetime has been extended to arbitrary elliptical orbits in the exterior field of a slowly rotating mass using the weak-field and slow-motion approximation scheme of GTR in~\cite{Mashhoon:2001ka}. The generalization of GCE to timelike world lines in Kerr spacetime is contained in~\cite{Hackmann:2014aga}.

Next, imagine the GCE for  spinning  test particles in orbit about the central rotating mass. This case has been investigated by a number of authors~\cite{Bini:2004md, Bini:2004me, Mashhoon:2006fj, F1, F2, Shahrear:2007zz, FaSh, EsFa}.

It turns out that the GEM approach can be extended to nonlocal gravity theory and the GCE has been considered within this nonlocal framework. The magnitude of the nonlocal contribution to GCE for the Earth is found to be smaller than about $10^{-10}$ of the GTR value due to the existence of galactic length scales in nonlocal gravity theory~\cite{Mashhoon:2019jkq}.


\subsection{GCE in the Presence of  Cosmological Constant $\Lambda$}

We wish to extend the discussion of GCE in Section II by including the cosmological constant $\Lambda$. We follow the treatment given in~\cite{Kerr:2003bp}. The cosmological constant plays the role of dark energy in current cosmological models for the accelerated expansion of the universe. We assume, as in current cosmological models, that $\Lambda$ is positive and $\Lambda \approx 10^{-56}$ cm$^{-2}$. For the rotating gravitational source, we employ the Kerr--de Sitter metric~\cite{Dem, Carter, KdS011} given by
\begin{align}\label{La}
\nonumber \left(ds\right)_{\rm KdS}^2={}&- \left[ 1- \frac{2Mr}{\Sigma}- \frac{\Lambda}{3} (r^2 + a^2 \sin^2\theta)\right] \left(dt\right)^2 - 2 a \left[\frac{2Mr}{\Sigma} + \frac{\Lambda}{3} (r^2 + a^2)\right] \sin^2\theta\, dt \,d\varphi \\
{}& +\frac{\Sigma}{\tilde{\Delta}}\left(dr\right)^2+ \frac{\Sigma}{\chi}\, \left(d\theta\right)^2 + \left[ \frac{2Mr}{\Sigma}\,a^2 \sin^2\theta + \left(1+\frac{\Lambda}{3} a^2\right) (r^2+a^2)\right]\,\sin^2\theta\, (d\varphi)^2\,,
\end{align}
where $\Sigma$ and $\Delta$ are the same as in the Kerr metric~\eqref{I4}, while
\begin{equation}\label{Lb}
\tilde{\Delta} := \Delta - \frac{\Lambda}{3} (r^2 + a^2)r^2\,, \qquad \chi := 1 + \frac{\Lambda}{3} a^2\,\cos^2\theta\,.
\end{equation}

For circular motion of test particles in the equatorial plane of the Kerr--de Sitter source, Eqs.~\eqref{I9}--\eqref{I11} for coordinate periods hold, except that the Keplerian frequency is now modified by the presence of the cosmological constant; that is, $\omega_0 \to \tilde{\omega}_0$, where
\begin{equation}\label{L1}
\tilde{\omega}_0^2 = \frac{M}{r_0^3} - \frac{1}{3} \Lambda\,.
\end{equation}
Moreover, stable circular orbits exist in this case out to radius $(3M/\Lambda)^{1/3}$, where $\tilde{\omega}_0$ vanishes~\cite{Howes}. As before, we find in this case,
\begin{equation}\label{L2}
 \tilde{t}_{+} - \tilde{t}_{-} =  4 \pi a\,;
\end{equation}
therefore,  according to the proper time of the preferred observer that is spatially at rest at $r_0$, the proper temporal difference for the two free counter-revolving test particles to complete the orbit of radius $r_0$ is given by
\begin{equation}\label{L3}
\tilde{\tau}_{{\rm S} +} - \tilde{\tau}_{{\rm S} -} = 4 \pi a \left[1- \frac{2M}{r_0} -\frac{\Lambda}{3} (r_0^2 + a^2)\right]^{1/2}\,.
\end{equation}
For counter-revolving spaceborne clocks, the analogue of Equation~\eqref{I16} is in this case
\begin{equation}\label{L4}
\tilde{\tau}_{\pm} = \frac{2\pi}{\tilde{\omega}_0} \left(1 -3\frac{M}{r_0}  -\frac{\Lambda}{3} a^2 \pm\, 2 a \tilde{\omega}_0\right)^{1/2}\,.
\end{equation}
Hence,
\begin{equation}\label{L5}
\tilde{\tau}_{+} - \tilde{\tau}_{-} =  \frac{4\pi a}{\left(1 -3\frac{M}{r_0}\right)^{1/2}}\left[ 1 + \frac{Ma^2 (1-\Lambda r_0^2)}{2r_0(r_0-3M)^2} + O(\epsilon^4)\right]\,,
\end{equation}
where $\epsilon = a/M$. We conclude that the GCE is at the lowest order independent of the cosmological constant. In fact, far from the source, i.e. $3M << r_0 << (3M/\Lambda)^{1/3}$, GCE basically depends only upon $a = J/M$, as before.


\subsection{GCE in the Hartle--Thorne Spacetime}\label{subsec:HT}

Next, let us extend the treatment of GCE to the exterior gravitational field of a slowly rotating relativistic star with quadrupole moment $q$. In the absence of rotation, the exterior field of the star is the static spherically symmetric Schwarzschild solution of mass $M$. The Hartle--Thorne solution is a perturbed Schwarzschild solution with mass $M$, angular momentum $J$ and quadrupole moment\footnotemark \footnotetext{
For the quadrupole mass moment of astrophysical compact objects, see, e.g., \cite{Shiba98, LarPoi99, Ster03}, and references therein. To avoid confusion, we note that $q$ is sometimes used to denote a dimensionless parameter connected to the quadrupole mass moment of a body. As an example, according to \cite{LarPoi99}, it is defined as $q := c^4\,Q/\left(M^3\,G^2\right)$, where $Q$ is the object's quadrupole moment having the dimensions of  mass times squared length.
 } $q$  that is valid to second order in $J$ and first order in $q$~\cite{Hartle:1968si}. Some of the properties of this spacetime have been elucidated and its quasinormal modes have been analytically calculated in the eikonal limit via the light-ring method in~\cite{Allahyari:2018cmg}. The metric of the Hartle--Thorne solution is given by
\begin{equation}\label{Ha}
\left(ds\right)_{\rm HT}^2=- \mathbb{F}_1\left(dt\right)^2+\frac{1}{\mathbb{F}_2}\left(dr\right)^2+\mathbb{G} r^2\, \left[\left(d\theta\right)^2 +\sin^2\theta\,\left(d\varphi-\frac{2J}{r^3}\,dt\right)^2\right]\,,
\end{equation}
where
\begin{equation}\label{Hb}
\mathbb{F}_1 = \left(1-\frac{2M}{r} +  \frac{2J^2}{r^4}\right)\left[1 +  \frac{2J^2}{Mr^3}\left(1+\frac{M}{r}\right)P_2(y) + 2 q' Q_2^2(x)P_2(y)\right]\,,
\end{equation}
\begin{equation}\label{Hc}
\mathbb{F}_2 = \left(1-\frac{2M}{r} +  \frac{2J^2}{r^4}\right)\left[1 +  \frac{2J^2}{Mr^3}\left(1-5\,\frac{M}{r}\right)P_2(y) + 2 q' Q_2^2(x)P_2(y)\right]\,,
\end{equation}
\begin{equation}\label{Hd}
\mathbb{G} = 1 -  \frac{2J^2}{Mr^3}\left(1+ 2\frac{M}{r}\right)P_2(y) + 2 q' \left[\frac{2M Q_2^1(x)}{\sqrt{r(r-2M)}} - Q_2^2(x)\right]P_2(y)\,.
\end{equation}
Here, $x = -1 +r/M$, $y = \cos \theta$ and $q'$ is a dimensionless parameter related to the quadrupole moment $q$ by
\begin{equation}\label{He}
q' = \frac{5}{8} \frac{qM - J^2}{M^4}\,.
\end{equation}
Moreover, $P_n$ is the Legendre polynomial of degree $n$, $P_2 (\cos \theta) = (3\cos^2\theta - 1)/2$, and $Q_n^m(x)$ is the associated Legendre function of the second kind for $x \in [1, \infty)$; in particular, for $r \gg M$, $x \gg 1$,
\begin{equation}\label{Hf}
Q_2^1 \sim \frac{2M^3}{5r^3}\,, \qquad Q_2^2 \sim \frac{8M^3}{5r^3}\,.
\end{equation}
Regarding the connection between the Hartle--Thorne and Kerr metrics, we note that the Kerr quadrupole moment is given by $J^2/(Mc^2)$ such that $q' =0$ for the Kerr spacetime. With $q' = 0$ in Eqs.~\eqref{Ha}--\eqref{Hd}, a coordinate transformation  reduces the Hartle--Thorne metric to the Kerr metric when terms higher than second order in $J$  are neglected~\cite{Allahyari:2018cmg}.

Using Equation~(147) of~\cite{Allahyari:2018cmg}, we find after some algebra that the analogue of Equation~\eqref{I10} for equatorial circular orbits is in this case given by
\begin{equation}\label{H1}
\frac{dt}{d\varphi}  = \pm \frac{1}{\omega_0} \left( 1 \pm a \omega_0 + a^2 \omega_0^2 - \bar{q} - \frac{7}{54} \frac{J^2}{M^4}\right)\,,
\end{equation}
where
\begin{equation}\label{H2}
\bar{q}  = \frac{15}{32} (28 - 25 \ln 3) \frac{qM - J^2}{M^4}\,.
\end{equation}
 Integrating over the azimuthal angle $\varphi$ as before, we find
\begin{equation}\label{H3}
t_{\pm}  = \frac{2\pi}{\omega_0} \left(1 \pm a \omega_0 + a^2 \omega_0^2 - \bar{q} - \frac{7}{54} \frac{J^2}{M^4}\right)\,.
\end{equation}
Hence, we obtain to second order in $J$,
\begin{equation}\label{H4}
 t_{+} - t_{-} = 4 \pi a\,.
\end{equation}

It is straightforward to find $d\tau/d\varphi$ using the Hartle--Thorne metric and the corresponding results that are given in~\cite{Allahyari:2018cmg}. We obtain, after some algebra,
\begin{equation}\label{H5}
\frac{d\tau}{d\varphi} = \pm\,\frac{1}{\omega_0} \left(1 -3\frac{M}{r_0} \pm\, 2 a \omega_0  - 2\bar{q} - \frac{7}{27} \frac{J^2}{M^4}\right)^{1/2}\,,
\end{equation}
where in the expressions proportional to $q$ and $J^2$ we have used only the lowest-order terms in $M/r_0 \ll 1$ for the sake of simplicity. It follows that
\begin{equation}\label{H6}
\tau_{\pm} = \frac{2\pi}{\omega_0} \left[\left(1 -3\frac{M}{r_0} \right)^{1/2}- \bar{q} - \frac{7}{54} \frac{J^2}{M^4}\right]  \pm\, \frac{2\pi a}{\left(1 -3\frac{M}{r_0}\right)^{1/2}}\,.
\end{equation}
Therefore, up to terms proportional to $J^2$,
\begin{equation}\label{H7}
\tau_{+}  - \tau_{-} = \frac{4\pi a}{\left(1 -3\frac{M}{r_0} \right)^{1/2}}\,.
\end{equation}
For $M/r_0 \ll 1$, we obtain to second order in $J$, $\tau_{+}  - \tau_{-} \approx 4\pi a$, as before.

The quadrupole moment $q$ in the exterior Hartle--Thorne solution could be of rotational origin with $q \propto J^2$ due to the existence of the centrifugal force. Alternatively, the Hartle--Thorne solution could describe the exterior of  an oblate spheroid with intrinsic $q$ that experiences uniform rotation. Indeed, it is possible to set $J = 0$ in the exterior Hartle--Thorne solution and the result is  a static solution with mass $M$ and quadrupole moment $q$. The Einstein tensor in this case is of order $q^2$; hence, this static solution is valid to first order in $q$ and can undergo rigid rotation to become the stationary Hartle--Thorne solution. In these stationary axisymmetric systems that are symmetric about the equatorial plane, we find no contribution of $q$ or $J^2$ terms to the GCE. Moreover it is important to recognize that solutions of GTR with quadrupole and higher moments are not  in general unique.


\subsection{GCE for Rigidly-Rotating Oblate Spheroid}\label{subsec:oblate}

In the Hartle--Thorne solution, terms of order $q\,J$ and higher have been neglected in the spacetime metric; however, the exterior gravitomagnetic field of a rotating mass could in general contain such terms that would lead to contributions to the GCE due to spin octupole and higher spin moments. Indeed,  the exterior gravitomagnetic field of a uniformly rotating axisymmetric configuration with its center of mass at the origin of spatial coordinates has been worked out in the post-Newtonian (pN) approximation scheme and the contribution of spin octupole moment to the GCE has been pointed out in~\cite{Mashhoon:1999nr, PS14}. In GTR, the exterior gravitational field of a rotating mass in general contains both mass moments and spin moments; in fact, the existence of a mass quadrupole moment can lead to a spin octupole moment.

To explain the additional contributions to the rudimentary clock effect in this case, let us start with the static  exterior field of a uniform-density oblate spheroid expanded  to the first pN (i.e. 1pN) level. The exterior field is static; therefore, no GCE is expected in this case. The 1pN acceleration ${\boldsymbol{A}}_\mathrm{pN}^\mathrm{obl} = d\boldsymbol{v}/dt$ experienced by a test particle orbiting an extended body of mass $M$, equatorial radius $R_\mathrm{e}$ and dimensionless quadrupole moment $J_2$ is\footnotemark{} \footnotetext{The 1pN acceleration ${\boldsymbol{A}}_\mathrm{pN}^\mathrm{obl}$ was derived earlier in \cite{Sof88, Sof89, Brum91} for a particular orientation of the body's symmetry axis in space.} \cite{Will14}
\begin{equation}
{\boldsymbol{A}}_\mathrm{pN}^\mathrm{obl} = {\boldsymbol{A}}^\mathrm{I} + {\boldsymbol{A}}^\mathrm{II} + {\boldsymbol{A}}^\mathrm{III}\,,
\end{equation}
where
\begin{align}
{\boldsymbol{A}}^\mathrm{I} \label{A1} & = \frac{3 G M J_2 R_\mathrm{e}^2}{2 c^2 r^4}\left(v^2 - \frac{4 G M}{r}\right)\left[\left(5 \xi^2 - 1\right)\mathbf{\hat{r}} - 2 \xi \mathbf{\hat{J}}\right]\,, \\ \nonumber \\
{\boldsymbol{A}}^\mathrm{II} \label{A2} & = -\frac{6 G M J_2 R_\mathrm{e}^2}{c^2 r^4}\left[\left(5 \xi^2 - 1\right)v_r - 2 \xi \lambda\right]\boldsymbol{v}\,, \\ \nonumber \\
{\boldsymbol{A}}^\mathrm{III} \label{A3} & = -\frac{2 G^2 M^2 J_2 R_\mathrm{e}^2}{c^2 r^5}\left(3 \xi^2 -1\right)\mathbf{\hat{r}}\,.
\end{align}
Here,
\begin{align}
v_r = \boldsymbol{v}\cdot\mathbf{\hat{r}}\,, \qquad \xi := \mathbf{\hat{J}}\cdot\mathbf{\hat{r}}\,, \qquad  \lambda := \mathbf{\hat{J}}\cdot\boldsymbol{v}\,
\end{align}
and $\mathbf{\hat{J}}$ is the unit vector of the body's axis of symmetry. Moreover, $J_2$ is the dimensionless mass quadrupole moment~\cite{Capde05}; indeed, for the constant density spheroid under consideration here we have
\begin{equation}
J_2 := \frac{q}{M\,R^2_\mathrm{e}} > 0\,.
\end{equation}

If the orbit lies in the body's equatorial plane, then $\xi=\lambda=0$. If, furthermore, the orbit is circular, i.e. if the radial velocity $v_r$ is zero, the acceleration in Equation\,\eqref{A2} vanishes, while the non-vanishing terms in Equation\,\eqref{A1} and Equation\,\eqref{A3}, which are entirely radial, are independent of the direction of motion of the test particle.

Let us now suppose that the  isolated  oblate spheroid of constant density  and ellipticity $\varepsilon := \sqrt{1 - R_\mathrm{p}^2/R_\mathrm{e}^2}$, where $R_\mathrm{p}$ is its polar radius, is rigidly and uniformly rotating. The gravitomagnetic spin-octupole\footnotemark{} \footnotetext{The spin-dipole term in \cite{PS14} yields the usual Lense--Thirring acceleration. For other studies on relativistic multipoles, see, e.g., \cite{multi1,multi2,multi3}.} acceleration ${\boldsymbol{A}}^\mathrm{oct}$ experienced by a test particle orbiting the spheroid at distance $r$ with velocity $\boldsymbol{v}$ is \cite{PS14}
\begin{equation}
{\boldsymbol{A}}^\mathrm{oct} = \frac{\boldsymbol{v}}{c^2}\boldsymbol{\times}{\boldsymbol{B}}^\mathrm{oct}\,,\label{eq1}
\end{equation}
where the gravitomagnetic field ${\boldsymbol{B}}^\mathrm{oct}$ can be calculated as \cite{PS14}
\begin{equation}
{\boldsymbol{B}}^\mathrm{oct} = -\nabla \phi^\mathrm{oct}\,,
\end{equation}
with the gravitomagnetic octupolar potential $\phi^\mathrm{oct}$ given by \cite{PS14}
\begin{equation}
\phi^\mathrm{oct} = \frac{6 G J R_\mathrm{e}^2\varepsilon^2}{7 r^4} P_3\left(\xi\right)\,.\label{eq3}
\end{equation}
Here, $P_3(\xi) = (5 \xi^3 - 3 \xi)/2$ is the Legendre polynomial of degree 3.
From Eqs.\,\eqref{eq1}--\eqref{eq3}, the spin-octupole  acceleration can finally be cast into the form
\begin{equation}
{\boldsymbol{A}}^\mathrm{oct} = \frac{3 G J R^2_\mathrm{e} \varepsilon^2}{7 c^2 r^5}\boldsymbol{v}\boldsymbol{\times}\left[ 5\xi\left(7\xi^2 - 3\right)\mathbf{\hat{r}} + 3\left(1-5\xi^2\right)\mathbf{\hat{J}}\right]\,.\label{eq4}
\end{equation}

If the orbital plane of the test particle coincides with the body's equatorial plane, i.e. if $\xi=0$, Equation\,\eqref{eq4} reduces to
\begin{equation}
{\boldsymbol{A}}^\mathrm{oct} = \frac{9 G R^2_\mathrm{e} \varepsilon^2}{7 c^2 r^5}\boldsymbol{v}\boldsymbol{\times}\mathbf{J}\,.\label{eq5}
\end{equation}
Also in this case, let us assume a circular orbit with radius $r_0$.
It can be noted that Equation\,\eqref{eq5} is directed radially inward or outward depending on whether the motion of the orbiter occurs in the opposite direction of the body's rotation or not; in the following, as before, the suffixes $\left(+\right)$ and $\left(-\right)$ appended to the particle's speed $v$ refer to the counterclockwise and clockwise directions of motion, respectively.
Thus, one has
\begin{equation}
\frac{v_{\pm}^2}{r_0} = \frac{GM}{r_0^2}\mp \frac{9 G J R_\mathrm{e}^2\varepsilon^2 v_{\pm}}{7 c^2 r_0^5}\,,\label{eq6}
\end{equation}
where the linear speed is related to the orbital angular speed $\omega_{\pm}$ by $v_{\pm} =\omega_{\pm}\, r_0 > 0$.
Therefore, Equation\,\eqref{eq6} can be written as
\begin{equation}
\omega^2_{\pm} \pm \Omega_\mathrm{oct}\,\omega_{\pm} -\omega_0^2 = 0\,,\label{eq7}
\end{equation}
where
\begin{equation}
\Omega_\mathrm{oct} := \frac{9 G J R_\mathrm{e}^2 \varepsilon^2}{7 c^2 r_0^5}\,.
\end{equation}
Assuming $\Omega_\mathrm{oct}\ll \omega_0$, Equation\,\eqref{eq7} implies
\begin{equation}
\omega_{\pm}\simeq \omega_0\left(1 \mp \frac{\Omega_\mathrm{oct}}{2\omega_0}\right)\,.\label{eq8}
\end{equation}
Using the definition
\begin{equation}
\omega_{\pm} := \frac{d\varphi_{\pm}}{dt}\, \label{eq9}
\end{equation}
in Equation\,\eqref{eq8} results in
\begin{equation}
d t \simeq \frac{d \varphi_{\pm}}{\omega_0}\left(1 \pm \frac{\Omega_\mathrm{oct}}{2\omega_0}\right)\,.\label{eq10}
\end{equation}
By integrating Equation\,\eqref{eq10} from 0 to $2\pi$ for the counterclockwise orbit and from $-2\pi$ to 0 for the
clockwise one, we find the orbital periods for the opposite directions of motion
\begin{equation}
t^\mathrm{oct}_{\pm} = \frac{2\pi}{\omega_0}\left(1 \pm \frac{\Omega_\mathrm{oct}}{2\omega_0}\right) = T_0 \pm \frac{\pi\Omega_\mathrm{oct}}{\omega^2_0} = T_0\pm \frac{9 \pi J R_\mathrm{e}^2 \varepsilon^2}{7 c^2 M r_0^2}\,.\label{toct}
\end{equation}
From Equation\,\eqref{toct}, the following gravitomagnetic spin-octupole clock effect
\begin{equation}
t^\mathrm{oct}_{+} - t_{-}^\mathrm{oct} = \frac{18 \pi J R_\mathrm{e}^2 \varepsilon^2}{7 c^2 M r_0^2}\,\label{Dtoct}
\end{equation}
is obtained, which does depend on the orbital radius as $r_0^{-2}$. The net result is then
\begin{equation}
t_{+} - t_{-} = 4 \pi \frac{J}{Mc^2}\left[1 + \frac{9}{14} \frac{R_\mathrm{e}^2 - R_\mathrm{p}^2}{r_0^2} + \cdots\right]\,,
\end{equation}
\textcolor{black}{which can be expressed in an invariant form in terms of the proper time of the preferred observer located in the equatorial plane at $r_0$. } For the analogue of Equation~\eqref{I18} in this case, namely the GCE in terms of the proper times of spaceborne clocks, see~\cite{Mashhoon:1999nr, PS14}. \textcolor{black}{Regarding the near-Earth circular equatorial orbits, for instance, the extra oblateness correction is $\leq 4 \times 10^{-3}$, since $(R_\mathrm{e}^2 - R_\mathrm{p}^2)/R_\mathrm{e}^2 \simeq 6.57 \times 10^{-3}$ for the Earth.}


\section{Observational Aspects of GCE}\label{obse}

The perspectives of measuring the  simple GCE associated with Equation~\eqref{I11}, namely, $\tau_{{\rm S} +} - \tau_{{\rm S} -} \approx 4 \pi J/ (Mc^2)$, have been investigated in several works; see
\cite{Gron97, Mashhoon:1998nn, Tar00, Tarta, Lic00, Iorio01a, Iorio01b, Iorioetal05, Lichetal06}.

A potentially favorable scenario around Jupiter is preliminarily outlined in \cite{Tarta,Tar00} pointing out that
the clocks to be compared should be on board orbiting spacecraft and should be appropriately stable for sufficiently long times.
The angular momentum of Jupiter is as large as $J_\mathrm{Jup}\simeq 6.9\times 10^{38}\,\mathrm{kg\,m^2\,s^{-1}}$ \cite{Sof03}, that is $\simeq 10^5$ times larger than that of the Earth. On the other hand, the  ratio of the mass of the Earth to that of Jupiter is $\simeq 1/318$; thus, the Jovian GCE (i.e. $4 \pi a_\mathrm{Jup}$) is $\simeq 370$ times  larger than the terrestrial one, amounting to about $5 \times10^{-5}$ s.  This circumstance explains why, at least in principle, Jupiter is the most viable scenario in our solar system for the detection of GCE.

In a space-based dedicated experiment with artificial satellites orbiting the Earth, several  effects of gravitational and non-gravitational origin have to be carefully modeled since they may induce competing signatures acting as sources of major systematic bias. They are mainly listed in \cite{Lic00}, where a couple of laser-ranged passive satellites of LAGEOS-type flying at an height of about 7000 km are considered.
In \cite{Iorio01a}, the gravitational perturbations have been investigated, while \cite{Iorio01b} is devoted to the impact of the non-gravitational disturbing accelerations. A comprehensive analysis was performed in \cite{Iorioetal05}.
In \cite{Lichetal06}, the case of non-equatorial orbits is examined. Such a choice is motivated by the fact that most satellite laser ranging stations are located in the northern hemisphere; thus, inclined orbits to the Earth's equator are more suitable from an observational point of view.


\section{Discussion}\label{summ}

In its simplest version, the gravitomagnetic clock effect (GCE) involves free test particles or spaceborne clocks on geodesic circular equatorial orbits about an axisymmetric rotating source of mass $M$ and angular momentum $J$; to go around the gravitational source, the orbiters take more (less) time by about $2\,\pi J/ (Mc^2)$ on prograde (retrograde) orbits. In this review, we have derived the GCE in its most basic form within the framework of general relativity and have elucidated its essential features. Various aspects of the effect have been discussed. The GCE has thus far been the subject of numerous studies and the results of these investigations have been briefly examined. The main impediments regarding the observation of the GCE have been described. In this connection, the advantages of a possible future mission to Jupiter have been pointed out.



\medskip

%


\end{document}